\documentclass[aps,prd,twocolumn,superscriptaddress,amsmath,showpacs]{revtex4}
\usepackage{graphicx}
\usepackage{dcolumn}
\usepackage{bm}
\usepackage{natbib}

\newcommand{\be}{\begin{equation}}
\newcommand{\ee}{\end{equation}}
\newcommand{\ba}{\begin{eqnarray}}
\newcommand{\ea}{\end{eqnarray}}

\newcommand{\LCDM}{$ \Lambda $CDM}

\newcommand{\mr}[1]{\mathrm{#1}}
\newcommand{\mtc}[1]{\mathcal{#1}}
\newcommand{\mtr}[1]{\mathrm{#1}}

\newcommand{\nad}{n_{\mr{ad}}}
\newcommand{\nadI}{n_{\mr{ad1}}}
\newcommand{\nadII}{n_{\mr{ad2}}}
\newcommand{\niso}{n_{\mr{iso}}}
\newcommand{\ncor}{n_{\mr{cor}}}

\newcommand{\abs}[1]{\vert #1 \vert}

\urldef\cosmomc\url{http://www.cosmologist.info/cosmomc}

\begin{document}

\title{Constraints on primordial isocurvature perturbations and spatial curvature\\ by Bayesian model selection}

\author {Jussi V\"aliviita}
\affiliation {Institute of Cosmology \& Gravitation, University
of Portsmouth, Portsmouth PO1 3FX, United Kingdom}
\affiliation{Institute of Theoretical Astrophysics, University of Oslo, P.O.
Box 1029 Blindern, N-0315 Oslo, Norway}

\author {Tommaso Giannantonio}
\affiliation{Argelander-Institut f\"ur Astronomie der Universit\"at
  Bonn, Auf dem H\"ugel 71, D-53121 Bonn, Germany}

\begin {abstract}
We present posterior likelihoods and Bayesian model selection analysis for generalized cosmological models
where the primordial perturbations include correlated adiabatic and cold dark matter isocurvature
components. We perform nested sampling with flat
and, for the first time, curved
 spatial geometries of the Universe,
using data from the cosmic microwave background (CMB) anisotropies,
the Union supernovae (SN)
sample and a combined measurement of the integrated Sachs--Wolfe (ISW) effect.
The CMB alone favors a $ 3\% $ (positively correlated)
isocurvature contribution  in both the flat and curved cases. 
The non-adiabatic contribution to the observed CMB temperature variance
is $0 < \alpha_T < 7\%$ at 98\% CL in the curved case.
In the flat case, combining the CMB with SN data artificially biases
the result towards the pure adiabatic \LCDM{} concordance model, whereas in the curved case the favored level of non-adiabaticity stays
at 3\% level with all combinations of data.
However, the ratio of Bayes factors, or $\Delta\ln(\mr{evidence})$, 
is more than 5 points in favor of the flat adiabatic \LCDM{} model,
which suggests that the inclusion of the 5 extra parameters of the curved isocurvature model
is not supported by the current data. The results are very sensitive to the second and
third acoustic peak regions in the CMB temperature angular power: therefore a careful calibration
of these data will be required before drawing decisive conclusions on the nature
of primordial perturbations. 
Finally, we point out that 
the odds for the  flat non-adiabatic model are 1:3 compared to the curved adiabatic model.
This may suggest that it is not much less motivated to
extend the concordance model
with 4 isocurvature degrees of freedom than it is to study the spatially curved adiabatic
model, though at the moment the model selection disfavors both of these models.

\end {abstract}


\maketitle

\section {Introduction} \label {sec:intro}

Observations of the cosmic microwave background (CMB) \cite{Nolta:2008ih} and distant
supernovae (SN) \cite{Kowalski:2008ez} have shaped the Lambda cold dark matter (\LCDM) standard
model of Cosmology. However,
being based on phenomenology, this model still needs better
understanding of some of its phases, in particular the
origin of the perturbations and the recent-time accelerated expansion of the Universe.

The paradigm of cosmic inflation is often assumed to describe the
early history of the Universe;
however, many different inflationary theories exist \cite{Lyth:1998xn} that
are still compatible with current data from the CMB and the large
scale structure (LSS) of the Universe, and it is therefore interesting
to look for ways to distinguish between them.

The primordial perturbations are usually believed to have formed from
quantum fluctuations in the early Universe, stretched
by inflation. There are two possible types of perturbations that can be
thus generated: single-field inflation can only produce adiabatic (isentropic)
modes of curvature perturbation $\mathcal R$, while multi--field models can
also generate isocurvature (entropy) perturbations $\mathcal S$ \cite{Polarski:1994rz}.
In Ref.~\cite{Bucher:1999re} four classes of isocurvature perturbations were
identified: the cold dark matter (CDM), baryon, neutrino density, and
neutrino velocity isocurvature modes. 
A generic perturbation can be composed as a linear combination of these ones 
and an adiabatic mode. However, it has turned out difficult to find physical
mechanisms to stimulate for example the neutrino velocity isocurvature mode.
In this paper we study a correlated mixture of adiabatic and CDM isocurvature
modes (later called the \emph{mixed model}), 
that is naturally generated in  multi--field inflationary models and
in curvaton(-like) models \cite{Lyth:2001nq,Enqvist:2001zp,Bartolo:2002vf,Moroi:2001ct,Ferrer:2004nv}.
Other scenarios that may generate observationally compatible isocurvature
include axions \cite{Beltran:2006sq,Hertzberg:2008wr}, dilaton
\cite{Copeland:1997nw} and ekpyrotic \cite{Koyama:2007mg,Notari:2002yc}
models, brane inflation \cite{Copeland:2006hv,Koyama:2001ct}, large scale
magnetic fields \cite{Tsagas:1999ft,Giovannini:2006gz}, cosmic strings and
other topological defects \cite{Battye:1998xe,Takahashi:2006yc}, whereas
an isocurvature mode in interacting dark energy models may grow
catastrophically \cite{Valiviita:2008iv,Majerotto:2009np,Valiviita:2009nu}.

Physically the inclusion of a CDM isocurvature mode is well motivated. It simply
means that initially, at a time $t_{\mr{rad}}$, deep in the radiation dominated era on
super-Hubble scales, the relative number densities of CDM and photons
are not spatially constant, and therefore the total entropy perturbation $\textstyle\mathcal{S}(t_{\mr{rad}},{\mathbf x})$
does not vanish everywhere. 
In addition to the amplitude (or indeed the variance) of $\mathcal R$ and $\mathcal S$
at $t_{\mr{rad}}$, other
important observables of the primordial perturbations are the tilt of
their power spectrum, parameterized by the spectral index $n$, and
their (non-)Gaussianity.

At later times and lower energies, the CMB is an almost perfectly
isotropic radiation that has been generated at the epoch of hydrogen
recombination. At even more recent
times, some additional effects can alter the CMB, such as the
integrated Sachs-Wolfe (ISW) effect \cite{Sachs:1967er}, which is
originated by the decay of the gravitational potentials. The most recent and complete ISW data have been obtained by
\cite {Giannantonio:2008zi} combining data from the CMB and six galaxy catalogs.

It is remarkable that the initial conditions of the CMB are set by the
final conditions of inflation. For this reason, we can distinguish
between different 
models of the early universe
 if we can show
which initial conditions agree best with the CMB (and ISW) observations.
In principle, the latter should be particularly useful
in constraining the mixed models, and in breaking
parameter degeneracies that remain after using the CMB and SN data.
This is mainly because
the SN consist of purely background data,
and hence the SN data can constrain
the isocurvature contribution only \emph{indirectly} by constraining
certain background parameters (such as $\Omega_\Lambda$) which
are degenerate with isocurvature \cite{KurkiSuonio:2004mn}.
In addition to constraining the background, the
ISW data probe directly the perturbation power spectrum. This is
affected by the CDM isocurvature mode, in particular if
its spectral index $\niso$ is relatively large as found for example
in \cite{Sollom:2009vd}.
Unfortunately, after employing in this paper for the first time the ISW data for constraining the mixed model,
we will find that the current ISW data are not accurate enough for this purpose, but still improve the constraints on the spatial curvature.

Observations from
the CMB and LSS indicate that the primordial perturbations were inflationary-like,
almost Gaussian, and mainly adiabatic with an almost scale-invariant power
spectrum.
After the first serious constraints on an uncorrelated mixed model
\cite{Enqvist:2000hp}, and after ruling out pure CDM isocurvature
\cite{Enqvist:2001fu}, various mixtures of the adiabatic and isocurvature
modes have been tested. 
In particular, since the release of the Wilkinson Microwave Anisotropy Probe
(WMAP) data, observational constraints have been obtained, e.g., by
\cite{Beltran:2005gr,Beltran:2005xd,Andrade:2005gw,Lazarides:2004we,KurkiSuonio:2004mn,Beltran:2004uv,Parkinson:2004yx,Moodley:2004nz,Ferrer:2004nv,Valiviita:2003tu,Valiviita:2003ty,Dunkley:2004sv,Andrade:2003xb,Gordon:2003hw,Crotty:2003rz,Peiris:2003ff,Bennett:2003bz} for WMAP1 \cite{WMAP1},
by \cite{Kawasaki:2007mb,Keskitalo:2006qv,Trotta:2006ww,Seljak:2006bg,Lewis:2006ma,Bean:2006qz} for WMAP3  \cite{WMAP3} and most recently
for WMAP5  \cite{WMAP5} by \cite{Sollom:2009vd},
 and by \cite{Hamann:2009yf} in the particular case of axion (uncorrelated, $\niso=1$).  
In \cite{Sollom:2009vd} it is shown that the CDM isocurvature mode is not
required by a combination of current data \emph{if flatness is assumed}. Nevertheless,
 there were hints of a positively correlated 4\% contribution from the CDM
isocurvature mode \cite{Keskitalo:2006qv} in
WMAP3  at the $3\sigma$
level. The
``isocurvature feature'' in the 3-year data was identified to lie around
the second and third acoustic peaks of the CMB power spectrum, which changed significantly in the 5-year data due to new beam calibrations.

 The only existing work where a mixture of primordial adiabatic and isocurvature perturbations has been studied in non-flat case is \cite{Dunkley:2005va}. The focus there was in testing how much the (possible) presence of isocurvature modes affects the determination of the geometry (spatial curvature) of the Universe, based on WMAP1 data.

So our first
task is to assess what the \emph{current CMB data alone}
tell about the nature of primordial perturbations. Then we add other complementary data,
either SN or ISW or both of them, to see whether they tighten the constraints. 
Our approach differs from \cite{Sollom:2009vd}, where all data were directly combined, the ISW was not used, and flatness was imposed.

In order to quantify the preference of one model over another
we perform several computationally costly Bayesian evidence comparisons \cite{Bassett:2004wz},
taking as a reference model the spatially flat
adiabatic \LCDM{} model. 
In line with \cite{Sollom:2009vd},
 we employ the recently
developed MultiNest package \cite{Feroz:2008xx}.
In addition to several advantages, described in Appendix
\ref{sec:sampling}, compared to the conventional Monte Carlo Markov Chain
(MCMC) method,
MultiNest allows us, \emph{for the first time, to constrain correlated adiabatic and CDM
isocurvature initial perturbations also in spatially curved
universes},
 and in particular to calculate Bayesian evidences for these models.

We will test how much spatial curvature is allowed, and on the other hand, how much allowing
for the non-flat case changes the posterior likelihoods of other
parameters, for both the adiabatic and mixed models. Indeed we will show that assuming spatial flatness of the Universe
in isocurvature studies 
would strongly bias the results toward pure adiabaticity.

We perform the full evidence calculation for each combination of data sets. Although this is
computationally demanding, \emph{it is an imperative not to blindly combine all different
types of data sets into one big chunk} (in our case CMB\&SN\&ISW) without testing
what is the individual information gained from each of the data sets and whether the
data are consistent with each other. In particular, the
black-box method of rushing to combine all available data would be dangerous, if
there happened to be 'tension' between the data sets.
Then artificially tight constraints would follow, without a real physical meaning.
Therefore we strongly advocate the approach where we add the data sets to the analysis one by one.

As a side product of our analysis, we obtain a comprehensive comparison of
flat and curved \emph{adiabatic} \LCDM{} models too --- in the light of any combination of
the CMB and SN or ISW data. These results are complementary
to a recent work \cite{Vardanyan:2009ft} where baryon acoustic oscillation data were employed
along with the CMB shift parameter and SN data,
but the ISW data or the full CMB data were not used.

The plan of this paper is the following. After describing our chosen
parametrization for the initial conditions of perturbations in
Sec.~\ref{sec:theory}, we summarize the method of our analysis in Sec.~\ref{sec:method}.
Then we expose and comment on the results of the 
likelihood analysis in Sec.~\ref {sec:likelihoods}, report
the Bayesian evidences in Sec.~\ref{sec:evidences}, and
conclude in Sec.~\ref{sec:concl}.

\section {Primordial perturbations \& CMB} \label{sec:theory}

We parametrize the primordial perturbations
the same way as in \cite{KurkiSuonio:2004mn, Keskitalo:2006qv}.

The evolution of (scalar) fields during (multiple-field) inflation
generates a trajectory in field space.
Perturbations can be decomposed in  modes which are along the trajectory (curvature
perturbations) and normal to it (entropy perturbations) \cite{Gordon:2000hv,Gordon:2001ph}.
The history
of these perturbations at any scale $\lambda \sim k^{-1}$, where $k$ is the wave number (later referred to simply as
``scale''),
goes as follows: perturbations were generated during inflation (or by an alternative theory)
at a time $t_{\star} (k)$, when they were ``promoted'' from the
 quantum vacuum level to the classical level by horizon crossing. 
Then, they were super-horizon (i.e., super-Hubble) and  at
some point in the radiation era $t_{\mathrm{rad}}$ they acted as
seeds of the matter power spectrum $P(k)$ and CMB perturbations.

During the super-horizon evolution, the
perturbations $(\mathcal{R},  \mathcal{S})$ are not frozen, but
evolve from their original values $ (\mathcal{R_{\star}},  \mathcal{S_{\star}}) $ according to
\be \label{eq:transfer}
\begin{pmatrix} \mathcal{R}  \\ \mathcal{S} \end{pmatrix} =
\begin{pmatrix} 1 & T_{\mathcal{RS}} \\ 0 & T_{\mathcal{SS}} \end{pmatrix}
\begin{pmatrix} \mathcal{R_{\star}}  \\ \mathcal{S_{\star}} \end{pmatrix}.
\ee
The transfer functions $T_{XY} (t,k)$ describe the evolution of
the perturbations, and are generally model dependent. In this paper we
approximate them by power laws. It is important to highlight that the form
of Eq.~(\ref{eq:transfer}) means that in the absence of primordial
isocurvature perturbations, $\mathcal{S_{\star}} = 0$ implies that no
isocurvature modes will be created, and the adiabatic perturbations
remain constant.

By introducing explicit power laws for the transfer functions, and
defining a pivot scale $k_0$ and a relative scale
$\bar k \equiv k / k_0$, we can
write the auto-correlation and cross-correlation power spectra for the perturbations at $t_{\mr{rad}}$
\ba
\mathcal{P}_{\mathcal{R}} (k) &=& \mathcal{P}_{\mr{ad1}} + \mathcal{P}_{\mr{ad2}}\nonumber\\
& = & A_r^2 \bar k^{\nadI - 1} + A_s^2
\bar k^{\nadII - 1}  \nonumber \\
\mathcal{P}_{\mathcal{S}} (k) &=& \mathcal{P}_{\mr{iso}} = B^2 \bar k^{\niso - 1}  \label{eq:correlations} \\
\mathcal{C}_{\mathcal{RS}} (k) &=& A_s B \bar k^{n_{\mathrm{cor}} - 1}\nonumber.
\ea
The usual adiabatic case is recovered by setting $A_s = B
= 0$. Also note that here the correlated spectral index can be written
in terms of the others: $\ncor = (\niso + \nadII)/2$.

The angular power spectra of the CMB temperature ($TT$) and polarization ($EE$) auto-correlation,
the temperature-polarization ($TE$) cross-correlation, the matter-matter ($mm$) auto-correlation, and
the temperature-matter ($Tm$) cross-correlation (the ISW-LSS cross-correlation) 
which would follow with $A_r = A_s = B =1$,
are convolutions of the adiabatic and isocurvature transfer functions
$\Theta_{l,\mathcal{R}}^{(X)} (k), \Theta_{l,\mathcal{S}}^{(X)} (k)$ as follows
\ba
 \hat C_l^{XY\mathrm{ad1}} &=& 4\pi \int \frac{dk}{k} \left[ \Theta_{l,\mathcal{R}}^{(X)} (k)  \Theta_{l,\mathcal{R}}^{(Y)} (k) \right]
\bar
k^{n_{\mathrm{ad1}} - 1}  \\
 \hat C_l^{XY\mathrm{ad2}} &=&  4\pi \int \frac{dk}{k}  \left[ \Theta_{l,\mathcal{R}}^{(X)} (k)  \Theta_{l,\mathcal{R}}^{(Y)} (k) \right]
\bar
k^{n_{\mathrm{ad2}} - 1}  \\
 \hat C_l^{XY\mathrm{iso}} &=&  4\pi \int \frac{dk}{k}  \left[ \Theta_{l,\mathcal{S}}^{(X)} (k)  \Theta_{l,\mathcal{S}}^{(Y)} (k) \right]
\bar k^{n_{\mathrm{iso}} - 1} \\
 \hat C_l^{XY\mathrm{cor}} &=&  4\pi \int \frac{dk}{k} \left[ \Theta_{l,\mathcal{R}}^{(X)} (k) \Theta_{l,\mathcal{S}}^{(Y)} (k)\right.\nonumber\\ 
&& \left. \ \ \ \ \ \ \ \ \ \ \ + \Theta_{l,\mathcal{R}}^{(Y)} (k) \Theta_{l,\mathcal{S}}^{(X)} (k)
\right]
\bar
k^{n_{\mathrm{cor}} - 1},
\ea
where $X$ and $Y$ stand for either $T$, $E$, or $m$.
Via the above integrals, the transfer functions  
$\Theta_{l,\mathcal{R}}^{(X)} (k), \Theta_{l,\mathcal{S}}^{(X)} (k)$
relate the primordial perturbations at wave number $k$ to the
anisotropy at multipole $l$ today. These functions depend on all the
history of the Universe from the primordial time $t_{\mr{rad}}$ up to today,
and they can be calculated by a Boltzmann integrator, such as CAMB/Cosmomc \cite{CAMB,COSMOMC}, publicly available at \cosmomc.

The total angular power spectrum is a sum of the above contributions weighted
with the primordial amplitudes $A_r$, $A_s$, and $B$ at the pivot scale
\ba
C_l^{XY} & = & A_r^2  \hat C_l^{XY\mathrm{ad1}}  +  A_s^2  \hat C_l^{XY\mathrm{ad2}}\nonumber\\
&&  + B^2
  \hat C_l^{XY\mathrm{iso}} + A_s B  \hat C_l^{XY\mathrm{cor}}.
\ea
In order to have a parametrization more suitable to data analysis, we
redefine the amplitude parameters as 
\ba
A^2 &\equiv& A_r^2 + A_s^2 + B^2 \nonumber \\
\alpha &\equiv& \frac {B^2}{A^2} \nonumber \in [0,1] \label{eq:alphadef}\\
\gamma &\equiv& \mathrm {sign} (A_s B) \frac {A_s^2}{A_r^2 + A_s^2}
\in [-1 , 1],\nonumber
\ea
so that the total angular power spectra are
composed by adiabatic, isocurvature and correlated components as
\ba
C_l^{XY} &=& A^2 \Big[(1 \! - \! \alpha)(1 \! - \! |\gamma|) \hat C_l^{XY\mathrm {ad1}} + (1 \! - \!
\alpha) |\gamma| \hat C_l^{XY\mathrm {ad2}} \nonumber \\
&& +\alpha  \hat C_l^{XY\mathrm {iso}} + \mathrm{sign} (\gamma) \sqrt {\alpha (1
  - \alpha) |\gamma|} \hat C_l^{XY\mathrm {cor}} \Big] \nonumber \\
&\equiv & C_l^{XY\mathrm {ad1}} + C_l^{XY\mathrm {ad2}} + C_l^{XY\mathrm
  {iso}} + C_l^{XY\mathrm {cor}}.
\label{eq:total-Cl}
\ea
We can now constrain the amount of isocurvature modes for the
primordial perturbations by measuring the likelihood distributions of
the parameters $\alpha$ and $\gamma$. We call the above parametrization
of primordial perturbation spectra ``spectral index parametrization'' or
as a short-hand notation ``$n$-parametrization''. It has six parameters which
describe the primordial perturbations:
$\nadI$, $\nadII$, $\niso$, $A$, $\gamma$, and $\alpha$.

Additional derived parameters can be defined. 
For example, a parameter $\nad^{\mr{eff}}$
represents the spectral index for adiabatic modes obtained expressing
the adiabatic contribution as a single power law:
\begin{multline}
  n_{\mr{ad}}^{\mr{eff}}(\bar k) - 1 =
  \frac{d\ln{\cal P}_{\cal R}(\bar k)}{d\ln\bar k}\\
  = \frac{(\nadI \! - \! 1) (1 \! - \! |\gamma|) \bar k^{\nadI-1} +
    (\nadII \! - \! 1) |\gamma| \bar k^{\nadII-1}}{(1 \! - \! |\gamma|)\bar
    k^{\nadI-1} + |\gamma| \bar k^{\nadII-1}}\,.
\label{eqn:nadeffk}
\end{multline}
Our pivot-scale free measure of the non-adiabaticity will be
the total non-adiabatic contribution to the CMB temperature variance
\ba
\alpha_T &\equiv& \frac {\langle ( \delta T^{\mathrm{non-ad}})^2 \rangle} {\langle ( \delta T^{\mathrm{total}})^2 \rangle} \nonumber \\
&=& \frac {\sum_l (2l + 1) (C_l^{TT\mathrm{iso}} + C_l^{TT\mathrm{cor}})     } {\sum_l (2l + 1) C_l^{TT}}.
\label{eq:alphaT}
\ea

When some parameters of a model are not sufficiently tightly constrained by
the data, the posterior likelihood functions become sensitive to the assumed
prior probability densities for the parameters. Even when one assumes flat,
i.e., uniform, priors for the primary parameters of the model, the question
remains, which parameters are taken to be the primary parameters, since the
priors for the quantities derived from the primary parameters (derived
parameters) will not be flat. 
To avoid problems related to spectral indices
becoming unconstrained when the corresponding amplitude parameters have small
values, a parametrization in terms of amplitudes
at two different scales, $k_1$ and $k_2$, was proposed in \cite{KurkiSuonio:2004mn}, and employed in
\cite{Keskitalo:2006qv} and \cite{{MacTavish:2005yk}}. In this paper we use this ``amplitude parametrization''
as the basis of our analysis, but we show the final results in the
$n$-parametrization.

\begin{table*}
\begin{tabular}{|l|l|r|}
\hline
Parameter & Explanation & range (min, max)\\
\hline
 & \multicolumn{1}{c|}{Primary parameters} & \\
\hline
$\omega_b$ & physical baryon density; $\omega_b = h^2\Omega_b$ & (0.005, 0.100) \\
$\omega_c$ & physical cold dark matter density; $\omega_c = h^2\Omega_c$ &  (0.01, 0.99)\\
$\theta$ & sound horizon angle; $\theta = r_s(z_\ast)/D_A(z_\ast)$ & (0.5, 5.0) \\
 $\tau$ & optical depth to reionization & (0.01, 0.30) \\
$\Omega_K$ & curvature density; $\Omega_K = 1-\Omega_{\mr{tot}}$ & (-0.20, 0.10) \\
$\ln (10^{10} A_1^2)$ & $A_1^2$ is the overall primordial perturbation power at $k = k_1 = 0.002\,$Mpc$^{-1}$ &  (1.0, 7.0) \\
$\gamma_1$ & correlation amplitude at $k = k_1 = 0.002\,$Mpc$^{-1}$ & (-1.0, 1.0) \\
$\alpha_1$ & primordial isocurvature fraction at $k = k_1 = 0.002\,$Mpc$^{-1}$ &  (0, 1.0) \\
$\ln (10^{10} A_2^2)$ & $A_2^2$ is the overall primordial perturbation power at $k = k_2 = 0.05\,$Mpc$^{-1}$ & (1.0, 7.0)\\
$\gamma_2$ & correlation amplitude at $k = k_2 = 0.05\,$Mpc$^{-1}$ &  (0, 1.0) \\
$\alpha_2$ & primordial isocurvature fraction at $k = k_2 = 0.05\,$Mpc$^{-1}$ &  (0, 1.0)  \\
$A_{SZ}$ & amplitude of the SZ template for WMAP and ACBAR & (0, 2) \\
\hline
 & \multicolumn{1}{c|}{Derived parameters} & \\
\hline
$H_0$ & Hubble parameter [km/s/Mpc]; calculated from $\omega_b$, $\omega_c$, $\theta$, and $\Omega_K$ & tophat (40, 100) \\
$h$ & $h=H_0/(100\,\mbox{km/s/Mpc})$ & (0.40, 1.00) \\
$\Omega_m$ & matter density parameter; $\Omega_m = (\omega_b + \omega_c)/h^2$ & \\
$\Omega_\Lambda$ & vacuum energy density parameter; $\Omega_\Lambda = 1 - \Omega_K - \Omega_m$ & \\ 
$\nadI$ & spectral index of primordial uncorrelated adiabatic part; $\nadI - 1 = d\ln({\cal P}_{\mr{ad1}})/d\ln k$ & \\
$\nadII$ & spectral index of primordial correlated adiabatic part; $\nadII - 1 = d\ln({\cal P}_{\mr{ad2}})/d\ln k$ & \\
$\niso$ & spectral index of primordial isocurvature part; $\niso - 1 = d\ln({\cal P}_{\mr{iso}})/d\ln k$ & \\
$n_{\mathrm{ad}}^{\mathrm{eff}}$ & effective single adiabatic spectral index, Eq.~(\ref{eqn:nadeffk}) & \\
$\gamma$ & correlation amplitude at $k = k_0 = 0.01\,$Mpc$^{-1}$ & (-1.0, 1.0) \\
$\alpha$ &  primordial isocurvature fraction at $k = k_0 = 0.01\,$Mpc$^{-1}$ & (0, 1.0) \\
$\alpha_T $ & total non-adiabatic contribution to the CMB temperature variance, Eq.~(\ref{eq:alphaT})
&\\
\hline
\end{tabular}
\caption{Our primary nested sampling parameters and a selection of derived parameters. \label{tab:parameters}}
\end{table*}

The mapping from the amplitude parametrization to the spectral index
parametrization is easy to find from the definitions (\ref{eq:correlations})~and~(\ref{eq:alphadef}). The
spectral indices can be written in terms of the parameters of amplitude
parametrization as
\begin{eqnarray}
  \nadI - 1 & = & \frac{ \ln\left[\mtc{P}_{\mr{ad1}}(k_2) /
      \mtc{P}_{\mr{ad1}}(k_1)\right] }
  {\ln(k_2/k_1)}\, \label{eq:newnadI} \\
  \nadII - 1 & = &\frac{ \ln\left[\mtc{P}_{\mr{ad2}}(k_2) /
      \mtc{P}_{\mr{ad2}}(k_1)\right] }
  {\ln(k_2/k_1)}\, \label{eq:newnadII} \\
  \niso - 1 & = & \frac{ \ln\left[\mtc{P}_{\mr{iso}}(k_2) /
      \mtc{P}_{\mr{iso}}(k_1)\right] }
  {\ln(k_2/k_1)}\,,\label{eq:newniso}
\end{eqnarray}
where the first (uncorrelated) adiabatic, the second (correlated) adiabatic
and the isocurvature power at scales $k_i$ ($i=1,2$) are given by
\begin{eqnarray}
  \mtc{P}_{\mr{ad1}}(k_i) & = & A_i^2 (1 - \alpha_i) (1-\abs{\gamma_i})\, \\
  \mtc{P}_{\mr{ad2}}(k_i) & = & A_i^2 (1 - \alpha_i) \abs{\gamma_i}\, \\
  \mtc{P}_{\mr{iso}}(k_i)  &=  & A_i^2 \alpha_i\,,\label{eq:PS}
\end{eqnarray}
respectively. Then the amplitudes $A$, $\alpha$, and $\gamma$ at the pivot scale $k_0$
are obtained from the amplitude-parametrization amplitudes
$A_1$, $\alpha_1$, and $\gamma_1$ defined at $k_1$ by
\cite{KurkiSuonio:2004mn}
\begin{eqnarray}
  A^{2} & = & A_1^{2}\Big[(1-\alpha_1)(1-\abs{\gamma_1})\tilde{k}^{\nadI-1}\nonumber \\ 
&& \ \ \ +
  (1-\alpha_1)\abs{\gamma_1}\tilde{k}^{\nadII-1} +
  \alpha_1\tilde{k}^{\niso-1}\Big] \, \label{eq:newA} \\
  \alpha & = & 
{\alpha_1\tilde{k}^{\niso-1}} \times \Big[
  (1-\alpha_1)(1-\abs{\gamma_1})\tilde{k}^{\nadI-1} \nonumber \\ 
&& +
    (1-\alpha_1)\abs{\gamma_1}\tilde{k}^{\nadII-1} +
    \alpha_1\tilde{k}^{\niso-1} \Big]^{-1}\, \label{eq:newalpha} \\
  \gamma & = & \frac{\gamma_1\tilde{k}^{\nadII-1}}
  {(1-\abs{\gamma_1})\tilde{k}^{\nadI-1} +
    \abs{\gamma_1}\tilde{k}^{\nadII-1}}\,, \label{eq:newgamma}
\end{eqnarray}
where $\tilde{k} = k_0/k_1$, and the spectral indices are obtained from
Eqs.~(\ref{eq:newnadI}--\ref{eq:newniso}).
Since we assume that all the component spectra can be described by power laws,
$\gamma_1$ and $\gamma_2$ must have the same sign.  Hence, they are not
completely independent. To obtain independent primary parameters, we draw
$\gamma_1$ from the range $[-1,1]$, but $\gamma_2$ only from the range
$[0,1]$, and let $\gamma_1$ determine the sign of the correlation. 

Employing the mappings (\ref{eq:newnadI}--\ref{eq:newniso}) and
(\ref{eq:newA}--\ref{eq:newgamma}) we obtain the posterior likelihoods of
$\nadI$, $\nadII$, $\niso$, $A$, $\alpha$, $\gamma$ for a MultiNest run in the
amplitude parametrization (corresponding to flat priors for $A_1$, $\alpha_1$,
$\gamma_1$, $A_2$, $\alpha_2$, and $\gamma_2$).
However, if we want to convert the results obtained in the amplitude
parametrization to flat priors for the spectral indices, then the mappings
(\ref{eq:newnadI}--\ref{eq:newniso}) and (\ref{eq:newA}--\ref{eq:newgamma}) are not enough: we have to correct for the prior too. This
can be done by weighting the multiplicities in the MCMC chains (i.e. weighting
the posterior likelihood) by the Jacobian of the transformation (\ref{eq:newnadI}--\ref{eq:newniso}) and (\ref{eq:newA}--\ref{eq:newgamma}). If the original run was made using primary parameters $\{
\Theta_i \}$ (and flat priors for them), but we want to show the results with
flat priors for $\{ \tilde\Theta_i \}$, the multiplicities must be multiplied
by
\begin{equation}
  J = \left|\det\left(\frac{\partial\Theta_i}{\partial\tilde\Theta_j}\right)\right|\,.
  \label{eqn:jacobian}
\end{equation}

From a purely theoretical point of view one would naively think that the choice of
pivot scales $k_1$ and $k_2$ (or in the $n$-parametrization $k_0$) is only a matter of taste.
However, when performing the likelihood analysis and producing 1d or 2d  marginalized 
posterior likelihoods (or global Bayesian evidences), the integration weight changes dramatically
if we change the pivot scale. This is evident from Eq.~(\ref{eqn:jacobian}) above
and in figure 21 in Ref. \cite{KurkiSuonio:2004mn}, see $\niso$ in particular.
Indeed, the posterior constraints on all the parameters depend somewhat on the choice
of pivot scales, but the effect is strongest on poorly constrained parameters.

One should try to optimize the constraining power of the data, and in
\cite{Keskitalo:2006qv} it was shown that in $n$-parametrization the optimal choice
for $k_0$ is in the middle (in logarithmic sense) of the available data.
Hence one
should avoid choosing $k_0$ too close to the edges of the data. For example,
a common but unsuitable choice, $k_0 = 0.05\;$Mpc$^{-1}$ (most recently employed in
\cite{Sollom:2009vd}),  leads to very loose constraints on
$\niso$ (and hence on the other parameters), since at these scales the isocurvature
has hardly any effect on the CMB angular power. On the other hand, another
common choice, $k_0 = 0.002\;$Mpc$^{-1}$ is too close to the large scale end of the data. Our choice,
$k_0 = 0.01\;$Mpc$^{-1}$, matches with Refs.~\cite{KurkiSuonio:2004mn} and \cite{Keskitalo:2006qv}, and is also
supported by \cite{Cortes:2007ak}, where  $k_0 = 0.017\;$Mpc$^{-1}$
was found to lead to the most stringent constraints. Ref.~\cite{Cortes:2007ak} formulated the issue
by quantifying the center of the data to be the statistical center.
For the amplitude parametrization we choose  $k_1 = 0.002\;$Mpc$^{-1}$ and
$k_2 = 0.05\;$Mpc$^{-1}$ which allows an easy comparison to the other works.

In addition to avoiding problems with the poorly constrained spectral index $\niso$, the amplitude
parametrization has a great advantage when performing Bayesian evidence calculations.
The Bayesian evidence is sensitive to the chosen prior (ranges) of the parameters; see e.g. \cite{Trotta:2006ww}.
In the case of $n$-parametrization it is completely unclear what the ranges for spectral indices should be.
If one chose very small ranges for $\nadII$ and $\niso$, then a larger
evidence would follow than when allowing for wide ranges. 
This ambiguity and arbitrariness was recently faced in \cite{Sollom:2009vd}.
In the amplitude parametrization we avoid this problem, since the amplitudes have ranges from $-1$ or $0$ to $+1$
by definition.

\section {Details of the analysis} \label{sec:method}

We perform nested sampling likelihood analyses using a modified version
of the MultiNest
package \cite{Feroz:2008xx},
which is publicly available
at \href {http://www.mrao.cam.ac.uk/software/multinest} {http://www.mrao.cam.ac.uk/software/multinest}, and 
 which is a significantly more efficient
alternative to the
standard Monte Carlo Markov chain (MCMC) method. Since this method is fairly
new, we will describe its principles in Appendix \ref{sec:sampling}.

\subsection {Parameters}
\label{sec:parameters}

In Sec.~\ref{sec:theory} we discussed in detail how we parametrize the primordial
perturbations. In the conventional purely adiabatic case one needs an amplitude
and a spectral index, or amplitudes at two different scales, i.e. two parameters.
In our correlated adiabatic and isocurvature model we need these for
the uncorrelated adiabatic, for the correlated adiabatic, and for the isocurvature spectrum.
This makes up six perturbation parameters. In addition to these we have the conventional
background parameters which exist in both models. Their number is four in the case of a flat Universe
and five in the case of non-flat Universe. Finally, when comparing the models to the
CMB data we need the amplitude of the Sunyaev-Zel'dovich template $A_{SZ}$. 

Therefore the adiabatic \LCDM{} reference model has 7 (8) independent primary parameters while our extended
correlated isocurvature model has 11 (12) parameters in the case of flat (non-flat) Universe.
We summarize the parametrization and give the prior ranges, as well as list useful
derived parameters in Table~\ref{tab:parameters}.

We assume a flat (uniform) prior
over the given ranges for the primary parameters. The derived parameters will
have non-flat priors unless otherwise stated. Indeed, the top-hat prior on the
Hubble parameter $H_0$ introduces somewhat non-flat priors for $\omega_b$, $\omega_c$, $\theta$, and
$\Omega_K$, but this is irrelevant since over the peak of the posterior likelihood their
prior is sufficiently flat.
Importantly, also the resulting prior on $\alpha_T$ is rather flat as shown in \cite{Keskitalo:2006qv}.

\subsection {Data}
\label{sec:data}

The data we use are: the publicly available WMAP 5 years temperature and polarisation data (TT,
EE, TE) \cite{WMAP5} plus the smaller scale ACBAR data \cite{Reichardt:2008ay} for the CMB anisotropies, the
supernovae from the Union compilation \cite{Kowalski:2008ez} with systematic uncertainties
flag turned on, and the ISW data of the
cross--correlations between the CMB and six galaxy
catalogs by \cite{Giannantonio:2008zi}.

The ISW effect is a small secondary anisotropy which is added at
late times
onto the primary CMB anisotropies in case the Universe is undergoing a
transition to a curvature or dark energy phase. It is due to the decay
of the gravitational potentials while CMB photons are travelling
through potential wells and, as such, is correlated with the large
scale structure (LSS) of the Universe. Its small magnitude, about $10 \%$ of
the primary CMB, hinders a direct detection of its temperature power
spectrum, but the effect can be detected by cross--correlating the CMB
with some tracer of the LSS \cite{Crittenden:1995ak}.
This signal has been detected by many authors by cross-correlating
the WMAP CMB map with various galaxy catalogs (see references in \cite{Giannantonio:2008zi})
out to a median redshift of $z=1.5$. Most recently, this limit was extended by 
\cite{Xia:2009dr} using the latest quasars from the SDSS, improving the previous result
of \cite{Giannantonio:2006du}.
Detections of this effect have been used to constrain various aspects of cosmology
\cite{Gaztanaga:2004sk, Corasaniti:2005pq, Giannantonio:2008qr, Lombriser:2009xg, Giannantonio:2009gi}.

The data set we use for the ISW was obtained by
\cite{Giannantonio:2008zi} by cross--correlating the WMAP maps of the
CMB with six galaxy catalogs (2MASS, SDSS main galaxies, LRGs and
QSOs, NVSS, HEAO) in several bands of the electromagnetic
spectrum at median redshifts $0 < \bar z < 1.5 $. 
The data consist of 13 angular bins of the real space
cross--correlation functions (CCFs) between each catalog and the CMB, at
angles $ 0 \deg \le \vartheta \le 12 \deg$, thus having 78 data points,
whose off-diagonal covariance matrix is important.

For each sampled model, we calculate the theoretical cross power
spectra between temperature and matter perturbations, $C_l^{Tm}$, from a full
Boltzmann integration within CAMB, adding the relevant smoothing beams,
and then perform a Legendre transformation 
to obtain the CCF at the same angles as the
measurements.
Finally we ensure that the theoretical models are compatible with the
observed auto--correlation functions of the
catalogs by allowing the galactic bias parameter 
(the ratio between matter power and observed galaxy power, which is assumed constant for each catalog),
to vary for each model. 
We can then calculate the likelihood of each model
given the ISW data, assuming that the errors are Gaussian.

We chose not to use the measurements of the matter power spectrum
because the current data and theories do not describe accurately the mapping from
the redshift space luminous galaxy observations to the Fourier space ($k$-space)
galaxy power spectrum, and further to the underlying $k$-space matter power spectrum.
Therefore the \emph{shape} of the matter power spectrum is still under investigation \cite{WP}. 
As the isocurvature contribution modifies the shape and tilt of the
matter power spectrum, once the shape of the observational
matter power spectrum becomes well understood it will improve
constraints on $\niso$, in particular. It should be noted that
our model with a free $\niso$ differs from Ref.~\cite{Trotta:2002iz} where the adiabatic and isocurvature
compnents share the same spectral index. As the CMB data prefers predominantly adiabatic nearly scale invariant
perturbations, the common spectral index is forced near to 1, which in \cite{Trotta:2002iz} leads to
a conclusion that the isocurvature would not affect the matter power spectrum.

\section {Posterior likelihoods} \label{sec:likelihoods}

\begin {figure*}[t]
\includegraphics[width=0.47\linewidth, angle=0]{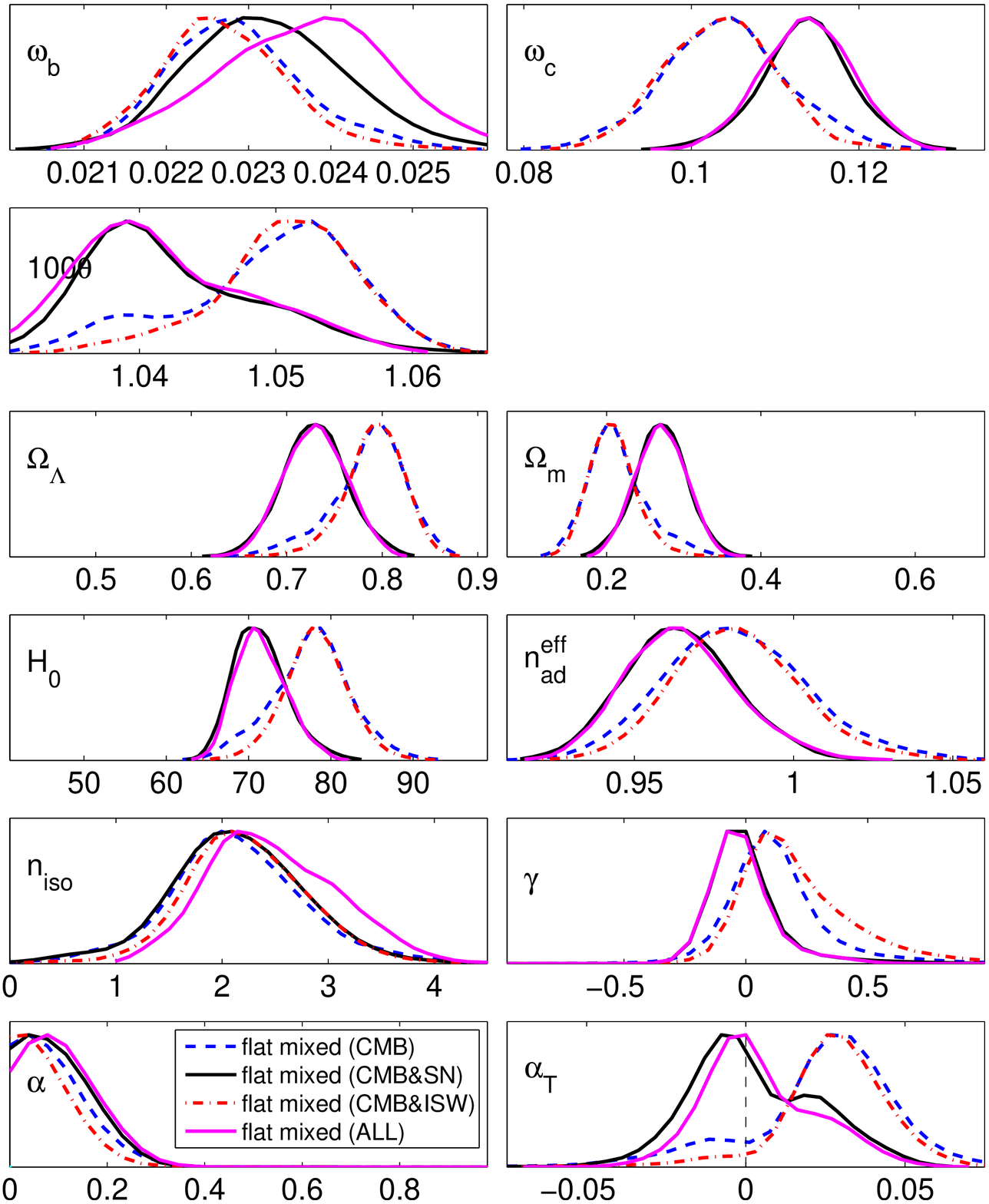}
\hspace{0.5cm}
\includegraphics[width=0.47\linewidth, angle=0]{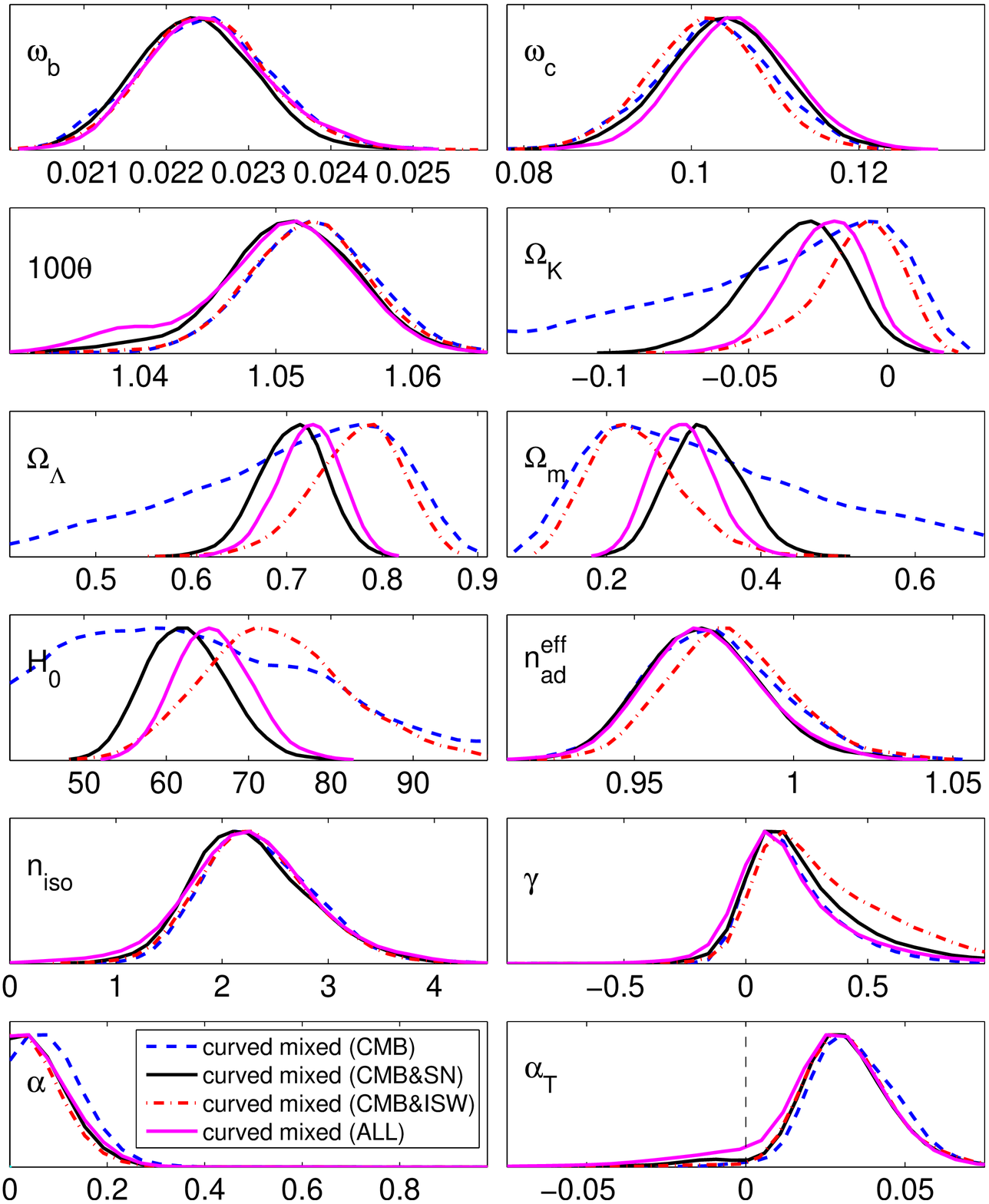}
\caption{\textbf{Left:} Posterior likelihood distributions for the model parameters assuming mixed initial conditions and  flat spatial geometry of the Universe. \textbf{Right:} 
The same for curved spatial geometry. 'ALL' refers to the CMB\&SN\&ISW data.}
\label{fig:flat}
\end{figure*}

We study the likelihoods with various combinations of data, comparing the results
 to the adiabatic \LCDM{} model. First we use the CMB data alone,
then we add either SN or ISW, and finally both of them into the analysis.
We present 1d marginalised posterior likelihoods in the mixed model
for the selected primary and derived parameters
in Fig.~\ref{fig:flat} for the flat (left panel) and
curved (right panel) cases.

\subsection{The CMB data alone}

The CMB alone does favor a small amount of positively correlated isocurvature mode.
This is consistent with what was previously reported in 
\cite{Keskitalo:2006qv}, although the few percent isocurvature contribution
is now slightly less favored (over pure adiabatic model, $\alpha_T=0$)
due to the modified shape of the 2nd and 3rd acoustic peaks in the WMAP5 data.

The key points in Fig.~\ref{fig:flat}
for the CMB data alone (blue dashed curves) are:
mixed models with a small contribution from a CDM isocurvature mode,
a small $\Omega_m$, a large $\Omega_\Lambda$, a large Hubble parameter $H_0$,
and a large sound horizon angle are marginally favored over
the concordance adiabatic \LCDM{} model. The CMB favors a positive
correlation, $\gamma$, between the primordial adiabatic and isocurvature perturbations
(with the sign convention where a positive primordial correlation
leads to a positive $C_l^{TT\mr{cor}}$ in the Sachs-Wolfe region, see
e.g. \cite{KurkiSuonio:2004mn,Keskitalo:2006qv}). 
In the curved case a positive non-adiabatic contribution
to the observed CMB temperature variance is more clearly favored than in the
flat case. 
We find that
$\alpha_T>0$ at 98.9\% (84.4\%) CL 
in the curved (flat) case, or
$0.8\% < \alpha_T < 6.5\%$ ($-3.1\% < \alpha_T < 6.7\%$ )
at 95\% CL; 
see the bottom right plots in Fig.~\ref{fig:flat}.

However, it is clear that due
to the poorly constrained
Hubble parameter, matter density, and curvature, we cannot use the current
CMB data alone for studying the mixed model. Nevertheless,
as we will discover in the next subsections, for a robust analysis
it is crucial to know what are the favored regions in parameter space with
the CMB data alone.

\begin {figure}[t]
\includegraphics[width=\linewidth, angle=0]{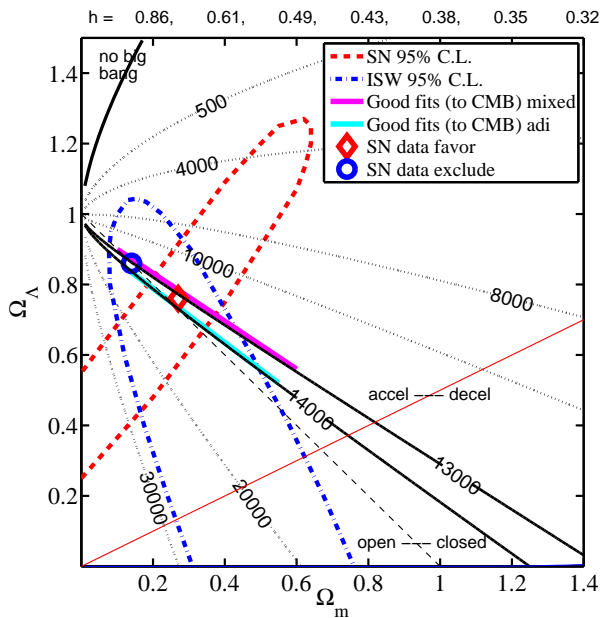}
\caption{Illustrative 95\% CL regions when using the SN or ISW data alone.
The dotted black curves indicate the angular diameter distance $D_A$ to last scattering
in units of Mpc. Two of them are highlighted --- in the mixed models
the CMB data favor $D_A \simeq 13\, 000\,$Mpc, whereas in the pure adiabatic models $D_A \simeq 14\, 000\,$Mpc 
is favored. The thin black dashed line indicates flat models. The thin red line indicates in which
part of the parameter space the expansion of the Universe is accelerating or decelerating today.}
\label{fig:jussithesis}
\end{figure}
\begin {figure} [t]
\includegraphics[width=0.95\linewidth, angle=0]{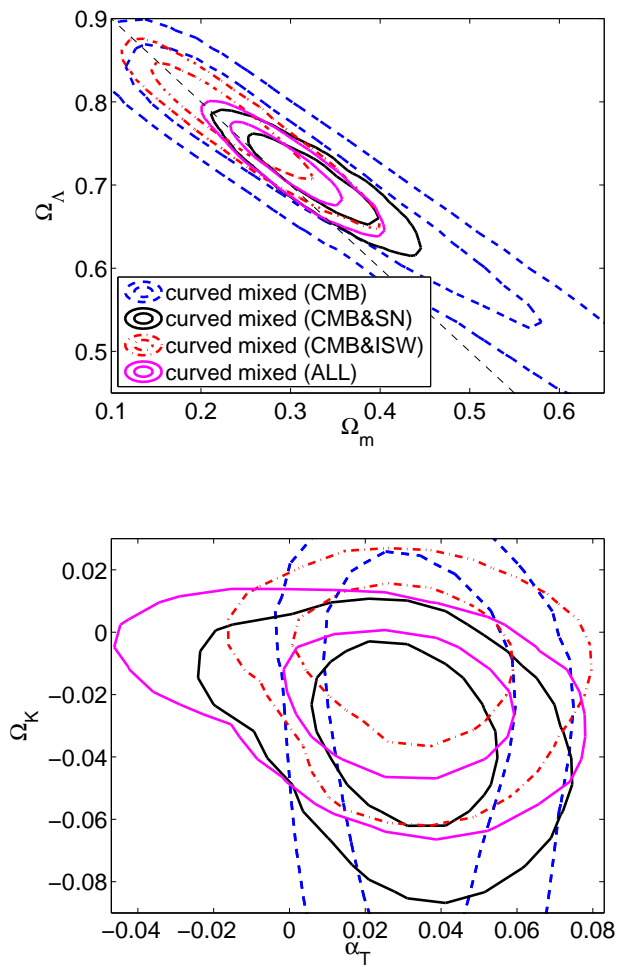}
\caption{Selected 2d posterior likelihoods for the mixed model when allowing for a spatially curved geometry of the Universe. The inner contours indicate 68\% CL, and the outer ones 95\% CL regions.}
\label{fig:2Dcurved}
\end{figure}

\subsection {Adding the SN and ISW data} 
\label{sec:SNetISW}

Now we repeat the likelihood analysis with the CMB and SN data. In the flat case the
data now prefer a purely adiabatic model: the likelihood of $\alpha_T$ has a peak
close to zero at slightly negative value. However, in the curved case adding the SN data hardly changes
any of the likelihood distributions of isocurvature parameters (compare blue
dashed and black solid curves in Fig.~\ref{fig:flat}).
Importantly the constraints of $\alpha_T$ remain almost the same as with the CMB data alone:
now $0 < \alpha_T < 7.0\%$ at 95\% CL.

The SN data do improve the constraints on some background parameters
($\omega_c$, $\Omega_K$, $\Omega_\Lambda$, $\Omega_m$), but do not significantly move
the peaks of their likelihoods, in the curved case. As noticed above,
the non-adiabatic contribution, $\alpha_T$, which is to some extent
degenerate with $\Omega_K$ and $\Omega_\Lambda$,  stays untouched.
We can understand this by looking at Fig.~\ref{fig:jussithesis}. 
The CMB tightly constrains the acoustic peak positions, and consequently the sound horizon angle
$\theta = r_s/D_A$, where $r_s$ is the sound horizon at last scattering and $D_A$ is the
angular diameter distance to last scattering. As $r_s$ depends only on $\omega_b$ and
$\omega_c$, and even this dependence is very mild \cite{Hu:2000ti}, the CMB constraint on $\theta$ is directly
reflected by the favored $D_A$, which depends on $\Omega_m$, $\Omega_\Lambda$ (or $\Omega_K$), and
$H_0$. The adiabatic model fits the acoustic peak positions perfectly whenever
$D_A \simeq 14\,000\,$Mpc (indicated by the highlighted cyan $D_A$ curve in Fig.~\ref{fig:jussithesis}).
However, in the mixed model there is an additional freedom
caused by the ability of the correlated contribution to the angular power spectrum,
$C_l^{TT\mathrm{cor}}$, to move slightly the acoustic
peak positions of the total $C_l^{TT}$ toward right, as shown in figure 2 in Ref.~\cite{Keskitalo:2006qv}.
Therefore a slightly larger $\theta$, i.e. a smaller $D_A$, is favored whenever there
is a small positively correlated isocurvature contribution to the CMB.
A $D_A\simeq  13\,000\,$Mpc (indicated by the highlighted magenta $D_A$ curve in Fig.~\ref{fig:jussithesis})
leads now to the best fits to the CMB. If we restrict the analysis to
flat models (indicated by the thin black dashed line), the CMB
picks the models (on the flat line, $\Omega_m + \Omega_\Lambda = 1$)
which are near the intersection point of the mentioned $D_A$ curve.
In the adiabatic case this means $\Omega_m \approx 0.26$ ($\Omega_\Lambda \approx 0.74$, $H_0\approx 72$),
whereas in the mixed case the intersection of $D_A\simeq  13\,000\,$Mpc and
the flat line is at  $\Omega_m \approx 0.19$ ($\Omega_\Lambda \approx 0.81$, $H_0\approx 87$).
This explains why, in light of the CMB alone, the flatness assumption forces
$\Omega_m$, ($\Omega_\Lambda$), and $H_0$ to quite unusual values, when
allowing for the mixed initial conditions of perturbations.

If we now combine the CMB with SN (indicated by red dashed 95\% CL curve in Fig.~\ref{fig:jussithesis}), it is
clear that the well-fitting flat mixed models (the blue circle) will be excluded. However,
in the curved case the SN data do not affect at all the well-fitting mixed models, and
the best-fit region stays unaffected (the red diamond symbol). Finally,
adding the ISW data (indicated by the blue dot-dashed 95\% CL curve in Fig.~\ref{fig:jussithesis})
does not affect at all the well-fitting adiabatic or mixed models. Therefore
the results for the isocurvature parameters with CMB\&ISW are very close
to the CMB alone case. The ISW data favor a slightly smaller matter density and slightly
less closed Universe than the SN data, thus affecting these background parameters
when compared to the CMB\&SN analysis.

All the above remarks are confirmed by the 2d posterior likelihood contours shown in the upper panel of Fig.~\ref{fig:2Dcurved}. The CMB alone
leaves a long degeneracy line in the ($\Omega_m,\Omega_\Lambda$) plane for mixed models with
$D_A\simeq  13\,000\,$Mpc. The SN or ISW data break this degeneracy, the
ISW data favoring slightly lower $\Omega_m$ than the SN data. Importantly, in the curved case, the well-fitting
models to the CMB sit in the middle of the best-fit region of SN or ISW. Combining all the data
(the magenta 68\% and 95\% CL curves) leads to the tightest constraints on $\Omega_m$ and
$\Omega_K$, being fully consistent with what we would expect from Fig.~\ref{fig:jussithesis}
and from the CMB\&SN and CMB\&ISW cases in  Fig.~\ref{fig:2Dcurved}.

The lower panel of Fig.~\ref{fig:2Dcurved} shows a tiny offset in the
favored curvature between the SN and ISW data, and indicates why
using all the data leads to looser constraints on $\alpha_T$, as seen
in the bottom right plot in the right panel of Fig.~\ref{fig:flat}.

The main conclusion after including the SN data is that, in the flat case,
this brings the result in line with the concordance adiabatic model, as also reported
in \cite{Sollom:2009vd}. Here we must put once more emphasis on the fact that this happens
only when restricting the analysis to flat models (thin black dashed lines in 
Figs.~\ref{fig:jussithesis} and \ref{fig:2Dcurved}), whereas when allowing
for spatial curvature of the Universe, a significant
non-adiabatic contribution remains allowed. Most importantly, the
well-fitting mixed models lie at the intersection of all the data
(CMB, SN and ISW) at slightly closed $\Omega_{\mr{tot}} \approx 1.03$ geometry,
though there is a slight competition (which exists also in the adiabatic case) between
the higher value of $\Omega_m$ preferred by the SN data, and the lower value
preferred by the ISW data.

\begin {figure}[t]
\includegraphics[width=\linewidth, angle=0]{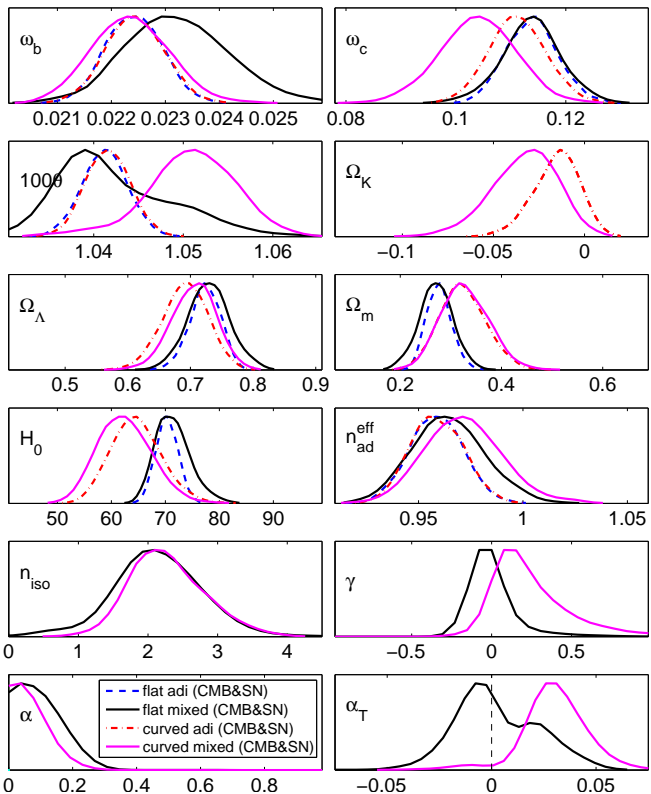}
\caption{Posterior likelihoods with the CMB\&SN data for selected model parameters
in the flat and curved case assuming either pure adiabatic or mixed initial conditions.
\label{fig:1dCMBetSNcomp}}
\end{figure}
\begin {figure}[t]
\vspace{-1mm}
\includegraphics[width=\linewidth, angle=0]{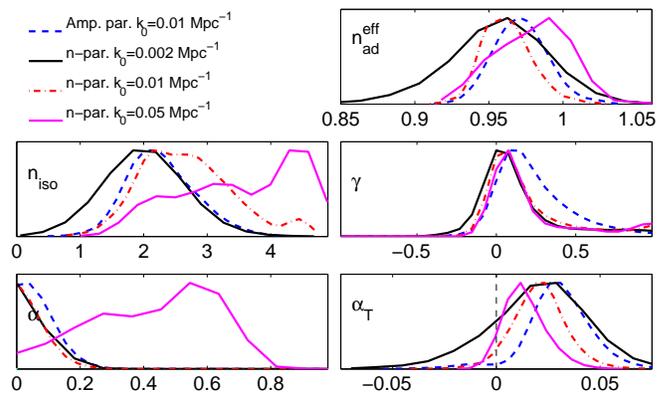}
\caption{Posterior likelihoods with the CMB\&SN data for selected model parameters
in the curved mixed model. 'Amp. par. $k_0=0.01$' indicates the results reported
in this paper, obtained assuming flat priors for the amplitudes $\alpha_{1,2}$
and $\gamma_{1,2}$ at scales $k_1=0.002\,$Mpc$^{-1}$ and
$k_2=0.05\,$Mpc$^{-1}$, and converted to spectral indices and amplitudes
at the pivot scale $k_0=0.01\,$Mpc$^{-1}$. 'n-par.\ $k_0=0.002$' indicates
what the results would be if we assumed flat priors in the spectral index
parametrization and chose the pivot scale $k_0=0.002\,$Mpc$^{-1}$.
'n-par.\ $k_0=0.01$'  and 'n-par.\ $k_0=0.05$' are  is the same as the previous
one, but choosing  $k_0=0.01\,$Mpc$^{-1}$ or $k_0=0.05\,$Mpc$^{-1}$,
respectively. The raggedness of the $k_0=0.05$ curves is due to a
small amount of well-fitting samples, after reweighting
by the Jacobian, Eq.~(\ref{eqn:jacobian}).}
\label{fig:pivot}
\end{figure}

\begin{table*}[t]
%
%
\begin{tabular}{|l|l|c|c|c|c|c|c|c|c|c|}
\hline
Data      & Model & $\chi^2$ & $\alpha_T$ & $\omega_c$ & $100\theta$ &
$\Omega_K$ & $\Omega_m$ & $H_0$ & $\nad^{\mr{eff}}$ & $\niso$ \\
\hline
CMB\&SN & flat adi     & 3003.9 & ---                         & 0.116 & 1.042  & ---                          & 0.285 & 69.8 & 0.956 \hfill (0.932,$\,$0.984) & --- \\
        & flat mixed   & 2999.0 & 0.002 \hfill (-0.03,$\,$0.05) & 0.111 & 1.037  & ---                          & 0.263 & 71.7 & 0.954 \hfill (0.930,$\,$1.000) & 3.5 \\
        & curved adi   & 3002.4 & ---                         & 0.107 & 1.042  & -0.03 \hfill (-0.04,$\,$0.01) & 0.370 & 59.2 & 0.948 \hfill (0.934,$\,$0.984) & --- \\
        & curved mixed & 2997.4 & 0.02 \hfill (0.00,$\,$0.07)  & 0.108 & 1.051  & -0.04 \hfill (-0.07,$\,$0.00) & 0.380 & 58.5 & 0.967 \hfill (0.934,$\,$1.009) & 2.9 \\
\hline
\end{tabular}
\caption{The best-fit $\chi^2$ and the best-fit values of selected parameters for
  the adiabatic and mixed models with the CMB\&SN data. In parenthesis, we indicate the minimal 95\% CL
  interval about the maximum of the corresponding 1d marginalized likelihood.}
\label{tab:bestfits}
\end{table*}

\subsection{Robustness of the main cosmological parameters against the assumed initial conditions}

An important question in constraining cosmologies is how much
the assumptions made in the analysis affect the interpreted values
of cosmological parameters from the given data \cite{Trotta:2001yw}. Often pure adiabatic
initial conditions are assumed when constraining the parameters of the \LCDM{} model.
In this subsection we show how the favored values (or regions) of the
main cosmological parameters change if one assumes mixed initial conditions.
In other words, we answer the question
``by assuming adiabaticity, would one find wrong constraints on the
main cosmological parameters, if the underlying 'true' initial perturbation
mode happened to be a correlated mixture of adiabatic and CDM isocurvature
perturbations?''.

Obviously, using the CMB data alone leads to rather large differences
between purely adiabatic and mixed models, in particular for
the posterior likelihoods of $\omega_c$, $\theta$, $\Omega_K$, $\Omega_\Lambda$, ($\Omega_m$),
$H_0$, the age of the Universe, and $\nad^{\mr{eff}}$. As it is unrealistic
to assume tight constraints in the mixed model with CMB data alone, we demonstrate
these effects in Fig.~\ref{fig:1dCMBetSNcomp} with CMB\&SN data.

In the flat case, the
pure adiabatic model favors smaller values of $\omega_b$, larger $\theta$, slightly smaller
$\Omega_\Lambda$ (hence larger $\Omega_m$), slightly smaller $H_0$, and
smaller  $\nad^{\mr{eff}}$. The CDM density $\omega_c$ remains unaffected.
Interestingly, the scale-invariant primordial adiabatic perturbations,
$\nad^{\mr{eff}}=1$, are within the 95\% CL region if mixed initial conditions are
assumed, while being far in the tail of the posterior likelihood if pure adiabaticity
is assumed.

In the curved case similar conclusions apply for $\Omega_\Lambda$ and $\nad^{\mr{eff}}$.
However, since now the SN data do not exclude a few percent positively correlated 
non-adiabatic contribution in the closed models (with $\Omega_{\mr{tot}} \simeq +1.03$) and
these models are actually slightly favored over the flat adiabatic models, larger
values of $\Omega_{\mr{tot}}$ will be preferred compared to the pure adiabatic case.
As explained in Sec.~\ref{sec:SNetISW}, the mixed model prefers a 
larger sound horizon angle, $\theta$. This, together with the curvature, affects in turn
$\omega_c$ and $H_0$.

\subsection{Best fits and 95\% CL intervals}

To complete the discussion about the posterior likelihoods, we report
in Table~\ref{tab:bestfits} the best-fit $\chi^2$ and selected best-fit 
parameters as well as 95\% CL intervals for some of these with the CMB\&SN
data (as seen in Fig.~\ref{fig:flat}, the CMB\&ISW or CMB\&SN\&ISW data lead
to very similar results).
The $\Delta\chi^2$ between the best-fit flat mixed and
flat adiabatic models is $-4.9$, while
the difference between the curved models is $-5.0$.
With the CMB alone these would be $-5.3$
and $-5.2$, respectively.
Noteworthy, the best-fit flat 'mixed' model is almost adiabatic, whereas
the best-fit curved mixed model has clearly non-zero isocurvature
contribution --- precisely as one would expect from the marginalized
likelihoods. 

As stated qualitatively in the previous subsection, the
determination of spatial curvature and the adiabatic spectral index
are significantly affected by the assumed initial conditions:
assuming mixed initial conditions, a more closed geometry
is favored than in the adiabatic case, and the flat geometry
is excluded at 95\% CL. On the other hand, the scale
invariant primordial adiabatic spectrum is excluded at much more
than 95\% CL if adiabatic initial conditions are assumed, whereas ---
not surprisingly, due to the extra freedom to modify the shape of the total initial
perturbation power spectrum --- in the mixed models the 95\% CL interval accommodates
the scale invariant spectrum. 
This result contradicts the claim in Ref.~\cite{Sollom:2009vd}
that the 'detection' of red adiabatic spectrum ($\nad^{\mr{eff}} < 1$)
would be robust against the inclusion of the CDM isocurvature mode.

The 95\% CL upper bound on the \emph{primordial contribution} of the
CDM isocurvature mode to the total perturbation power at the scale
$k_0 =0.01\,$Mpc$^{-1}$ is $\alpha < 22\%$ in the flat case, and
$\alpha<17\%$ in the curved case.

\subsection{Dependence on the pivot scale}

We do not report any constraints for $\niso$, since 
its posterior likelihood depends drastically on the chosen
parametrization, in particular on the choice of the pivot scale, as shown in Fig.~\ref{fig:pivot}.
While with some choices a scale invariant spectrum, $\niso=1$, is 'allowed',
in general fairly large values ($\niso\gtrsim 1.5$) seem to be favored.
There have not been many theoretical models that would predict such a large
isocurvature spectral index, but recently an explicit axion model, which leads
to $\niso \sim 2$---$4$, was constructed in \cite{Kasuya:2009up}.

The posterior likelihoods
obtained assuming flat priors in the amplitude parametrization
(which we use for reporting the results in this paper), agree
in general well with the more traditional spectral index parametrization,
where flat priors for spectral indices and amplitudes at a pivot scale
$k_0=0.01\,$Mpc$^{-1}$ are assumed. However, due to a different integration
measure upon marginalization, the posterior likelihoods for all
parameters differ from these, if we choose $k_0=0.002\,$Mpc$^{-1}$ or
$k_0=0.05\,$Mpc$^{-1}$, which are the most common choices in the literature.
In particular, the difference in the poorly constrained
$\niso$ is large. This effect was first realized in
\cite{KurkiSuonio:2004mn}, where it was strongly recommended that
in the isocurvature analysis one should adopt $k_0\simeq0.01\,$Mpc$^{-1}$,
which leads to the tightest constraints and minimizes the ambiguity (caused by
poorly constrained $\niso$)
in determining the main cosmological parameters. As a further improvement,
the amplitude parametrization, which we employ here, was suggested.
With a large pivot scale (small $k_0$), a small $\niso$ appears to
result in, whereas a small pivot scale (large $k_0$) leads to an apparent
peak of the likelihood at $\niso>4$. The most recent isocurvature analysis
\cite{Sollom:2009vd} suffers from this problem, since in \cite{Sollom:2009vd}
$k_0=0.05\,$Mpc$^{-1}$ is adopted.
From Fig.~\ref{fig:pivot} it is clear why \cite{Sollom:2009vd} reports very loose constraints on $\niso$,
and claims that very large spectral tilts seem to be favored.

\begin{table*}[ht]
\begin{tabular}{|l||c||c|c||c|c||c|c|}
\hline
Data & \multicolumn{7}{l|}{\hspace{5.9cm} Model} \\
             & \multicolumn{1}{c||}{Flat: adiabatic} & \multicolumn{2}{c||}{Non-flat: adiabatic} & \multicolumn{2}{c||}{Flat: mixed} & \multicolumn{2}{c|}{Non-flat: mixed}\\
            & $\ln(\mr{ev})$ &  $\ln(\mr{ev})$ & $\Delta \ln(\mr{ev})$ &  $\ln(\mr{ev})$ & $\Delta \ln(\mr{ev})$ &  $\ln(\mr{ev})$ & $\Delta \ln(\mr{ev})$ \\
\hline
CMB  \hfill    & $ -1370.3 \pm 0.3 $ & $ -1372.9 \pm 0.3 $ & $ -2.6 \pm 0.5 $ & $ -1375.5 \pm 0.4 $  &  $ -5.2 \pm 0.5 $ & $ -1376.5 \pm 0.4 $ & $ -6.2 \pm 0.5 $  \\
CMB\&SN \hfill      & $ -1525.7 \pm 0.2 $  & $ -1527.5 \pm 0.2 $ & $ -1.8 \pm 0.3 $ & $ -1531.3 \pm 0.3 $ & $ -5.6 \pm 0.4 $ & $ -1531.9 \pm 0.3 $ & $ -6.2 \pm 0.4 $ \\
\hspace{1.2mm} --- Sqrt param. \hfill     & as above  & as above & as above & $ -1528.6 \pm 0.2 $ & $ -2.9 \pm 0.3 $ & $ -1529.2 \pm 0.3 $ & $ -3.5 \pm 0.3 $ \\
CMB\&ISW \hfill    & $ -1393.7 \pm 0.3 $ & $ -1396.7 \pm 0.3 $  & $ -3.1 \pm 0.4 $  & $ -1397.2 \pm 0.3 $ & $ -3.6 \pm 0.4 $ & $ -1399.8 \pm 0.3 $  & $ -6.1 \pm 0.4 $ \\
CMB\&SN\&ISW \hfill& $ -1548.2 \pm 0.3 $  & $ -1551.6 \pm 0.3 $  &  $ -3.3 \pm 0.5 $  & $ -1553.7 \pm 0.4 $  & $ -5.5 \pm 0.5 $ &  $ -1555.6 \pm  0.3$  & $-7.4 \pm 0.4$ \\
\hline
\end{tabular}
\caption{Bayesian evidences for the flat and non-flat adiabatic and mixed (correlated adiabatic and isocurvature) models with various combinations of data. Columns $\ln(\mr{ev})$  stand for the natural logarithm of the evidence (total likelihood). Columns $\Delta\ln(\mr{ev})$ give the difference of $\ln(\mr{ev})$ of the considered model compared to the flat adiabatic model with the same combination of data; $\Delta\ln(\mr{ev}) = \ln({\mr{ev}}/\mr{ev}_{\mr{flat,adi}})$. Note that a negative  $\Delta\ln(\mr{ev})$ means that the model is disfavored compared to the flat adiabatic \LCDM{} model. 'Sqrt param.' refers to an alternative mixed model described in
Sec~\ref{sec:alternative}.\label{tab:evidences}}
\end{table*}

Apart from the issues with the spectral indices, our findings \emph{for the flat case} agree well with \cite{Sollom:2009vd} where
the recent CMB (WMAP5 \& ACBAR), SN (SNLS), and LSS (SDSS DR5 LRGs) data, and a Gaussian prior $\omega_b = 0.022 \pm 0.006$, were
employed in flat models, without testing the results with individual
combinations of the data, such as CMB\&SN or CMB\&LSS. Interestingly,
based on comparing our results with those of \cite{Sollom:2009vd},
the LSS data do not seem to improve the constraints on the
isocurvature. Moreover, the inclusion of LSS data is not enough to
overcome the unsuitable choice of pivot scale made in \cite{Sollom:2009vd}, although one would have expected the LSS to
improve the constraints on $\niso$.

\section{Bayesian evidences} \label{sec:evidences}

The main results of this paper, 
the Bayesian evidences (see e.g.\ Appendix \ref{sec:sampling} and Ref.~\cite{Trotta:2008qt}),
are presented in Table~\ref{tab:evidences}.
There we compare other models to the flat adiabatic \LCDM{} model, giving
$\Delta \ln(\mr{ev}) = \ln(\mr{ev}) - \ln(\mr{ev}_\mr{flat,adi})$.
Clearly the flat adiabatic \LCDM{} model is favored over curved adiabatic
model and over both flat and curved mixed models. In light
of the current data, the Bayesian model selection decisively ($\Delta \ln(\mr{ev}) < -5$)
disfavors curved correlated isocurvature model. This is because
the model selection punishes strongly for the 5 extra parameters
(compared to the adiabatic \LCDM{} model) whose inclusion does not
improve the fit to the data considerably:
see, e.g., the bes-fit $\chi^2$ values for CMB\&SN in Table~\ref{tab:bestfits}.
However, one should keep in mind
that in the mixed models with $\niso \gtrsim 2$, the main effect on the CMB
is a modified 2nd and 3rd acoustic peak region. Therefore the determination
of the isocurvature contribution is very sensitive to the calibration
of the CMB temperature angular power spectrum in this region.

\subsection{Evidences in an alternative parametrization}
\label{sec:alternative}

So far we have considered the primordial isocurvature perturbations in mixed models
parametrized by the amplitudes $\alpha_1$ and $\alpha_2$ ($\gamma_1$ and $\gamma_2$)
of the primordial isocurvature (correlation) power spectrum at two different scales.
It should be kept in mind that in Bayesian model selection,
'the model' means the theoretical set-up
\emph{including the chosen parametrization and the priors of these parameters}.
Therefore, by the very first principles of model selection, the evidences are
inevitably sensitive to the chosen parametrization. To account for this, we reproduce
a couple of our results for another mixed model, which is otherwise the same as the
previous model, but where the primordial non-adiabatic components are described by
amplitudes $\tilde\alpha_i$ ($\tilde\gamma_i$) of the primordial perturbations,
instead of the amplitudes of the power spectra. These two parametrizations are related by
\begin{equation}
\alpha_i = \tilde\alpha_i^2, \quad\quad
\gamma_1 = \mtr{sign}(\tilde\gamma_1)\tilde\gamma_1^2, \quad\quad
\gamma_2 = \tilde\gamma_2^2.
\end{equation}

While the posterior likelihoods of the other cosmological parameters remain almost
unchanged, and $\alpha_T$ is affected by much less than $1\sigma$, the
different integration measure affects considerably the global Bayesian evidence.
We show the evidences for our new mixed model, which we call 'sqrt parametrization' or 'sqrt model',
in Table~\ref{tab:evidences} for the CMB\&SN data. 
In light of the CMB\&SN data, the flat (curved) mixed
sqrt model is within $2.9$ ($3.5$) from the flat adiabatic \LCDM{} model,
corresponding to odds of 1:18 (1:33). Therefore,
there is strong --- but not decisive --- evidence against the sqrt model.
In particular,
taking into account the error estimates of $\Delta\ln(\mr{ev})$,
the flat mixed sqrt model is not
significantly disfavored when compared to the curved adiabatic \LCDM{} model.
As their evidence difference $-1.1$ corresponds to odds of 1:3, 
this suggests that in future studies, isocurvature should be
treated on a similar footing as checking for curved adiabatic models.

\section {Conclusions} \label {sec:concl}

In this paper we have presented a new likelihood and model selection analysis allowing 
for a correlated cold dark matter isocurvature mode of primordial perturbations,
\emph{for the first time including
spatial curvature.}

Taking first a frequentist's point of view,
we have shown in the light of posterior likelihoods
that models with a small fraction of isocurvature ($\simeq 3$\%)
are still favored by a CMB-only analysis, and including the Type Ia Supernovae or the integrated
Sachs-Wolfe effect data does not change
this result in the spatially curved Universe. 
Up to 7\% positively correlated non-adiabatic contribution is allowed at 95\% CL, whereas
the pure adiabatic model lies near the boundary of the 95\% CL region. 
In the flat case, we discover the previously known
result that the SN data cut out the best-fit isocurvature models. Interestingly this does
not happen in the curved case as indicated in Fig.~\ref{fig:2Dcurved}.
The ISW data constrain the vacuum energy density and
curvature of the Universe in a complementary way to CMB or SN. Therefore the inclusion of
the ISW data in the analysis sets more stringent constraints on the curvature, but
does not seem to tighten the constraints on isocurvature. 

We recommend including the
spatial curvature in the isocurvature analysis of future CMB data (combined with
some other probes of curvature, dark energy density, or Hubble parameter), since assuming
flatness of the Universe considerably --- and misleadingly --- biases the results toward pure adiabaticity.

The Bayesian model selection, which heavily penalizes for any ``unnecessary'' extra degrees of freedom,
disfavors strongly or decisively --- depending on the parametrization of primordial perturbations ---
the mixture of correlated adiabatic and isocurvature primordial perturbations,
\emph{in light of the current data}. However, one should keep in mind that this result is very sensitive to the calibration
of the CMB data around the second and third acoustic peaks, and therefore future data, e.g. from the Planck satellite, may
either weaken or strengthen the constraints.

From a theoretical point of view, other scenarios
that modify the second and third acoustic peak region
typically involve other types of isocurvature, such as a dynamical contribution from cosmic strings \cite{Bevis:2007gh}.
In addition, the kinematic Sunyaev-Zel'dovich effect modifies the same region \cite{GenovaSantos:2009tf}. However,
Planck should be able to distinguish between these and the mixed adibatic and CDM isocurvature model, since
the former produce just a single 'bump' of extra angular power, whereas the latter modifies the angular power spectrum in a more
complex way at a few percent level, as shown in figure 2 in Ref.~\cite{Keskitalo:2006qv}.   

Although we have focused on isocurvature, we have also constrained the
geometry of the Universe in the pure adiabatic model, finding that
with the CMB\&SN data the Bayesian model selection significantly (with odds
of 1:6) disfavors spatially curved adiabatic model compared
to the flat model. This constraint is tightened from 'significant' to
'strong' (with odds of 1:27) when we add the ISW data into the analysis.

It seems likely that the Bayesian model selection, with near-future
data, could decisively rule out both the spatially curved geometry of
the Universe and the mixed model, irrespectively of the
parametrization issues, while the frequentist's approach may continue
to slightly 'favor' these models over the flat \LCDM{{} model.

However, it should be noticed that even the future CMB \emph{temperature
anisotropy data alone} are unlikely to constrain the mixed models with
$\niso \lesssim 2$--$3$ better than the current WMAP data, as first
pointed out in \cite{jvPhDthesis} (compare also figures 2c and 13 in
\cite{KurkiSuonio:2004mn}, and see figure 1 in \cite{Hamann:2009yf}).
The reason is that on sub-horizon (sub-Hubble) scales before
last scattering, the CDM density perturbations resulting from the primordial
CDM isocurvature mode are damped by $k$ compared to those ones resulting
from the adiabatic primordial mode. (Note that in the power spectrum
this damping is $\propto k^2$, and in the angular power $\propto l^2$). Therefore
in order to significantly modify
the predominantly adiabatic angular power spectrum above the multipole
$l \gtrsim 200$, one needs a large isocurvature spectral index.
Consequently, models with $\niso\lesssim 2$ would modify the anisotropy spectrum
only at low multipoles, say $l \lesssim 200$, but here the temperature data
are already now cosmic-variance limited. In particular, the future
CMB \emph{temperature} data can not significantly improve the constraints on
a model where all the components share the same spectral index
($\niso =\nadI = \nadII = \ncor = n \approx 1$), such as in
\cite{Trotta:2002iz}. Nevertheless,  new accurate \emph{polarization}
data will help in reducing the uncertainty caused by the cosmic variance and
in breaking the parameter degeneracies. Polarization data together
with more accurate data on high multipoles also fix the background
parameters better, leading indirectly to tighter
constraints on isocurvature. As shown in \cite{Hamann:2009yf},
one can thus expect a moderate improvement on the constraints even for
models with nearly scale-invariant isocurvature spectrum.
Our main forecast, e.g., for Planck, is
that having more accurate data on high multipoles,
in particular helps to constrain the isocurvature spectral index in models
where it could be $\niso \gtrsim 2$ (in this paper more than 50\% of
the well-fitting mixed models, see Fig.~\ref{fig:flat}).
In addition, Planck, as well as future supernovae data,
will constrain the background parameters better and thus
\emph{indirectly} improve the constraints on the isocurvature contribution
by breaking the degeneracies.

The prospects of the ISW data are more pessimistic. We have checked with our
best-fit models that about 10 times more accurate ISW data would be
needed if one was to directly discern between the perturbation
spectra of the pure adiabatic and the mixed model, even with a large
$\niso$. Since the required accuracy is more than the theoretical
bound for the signal to noise \cite{Crittenden:1995ak}, the role of
the ISW data will remain limited to constraining the background, and
thus only \emph{indirectly} the isocurvature contribution.

\subsection*{Acknowledgments} 
We thank Alessandro Melchiorri for suggesting us to work on this subject, and
we acknowledge the work and communications with
Reijo Keskitalo at the early stages of this project.
We thank the Bavarian Academy of Science and the Leibniz Computer Center in Munich, Germany, for computational resources. 
In addition, a small part of the analysis was
performed at the Cosmos
supercomputer in Cambridge, United Kingdom. One of the cases
was run at the CSC, Finland. We thank the DEISA Consortium (\url{www.deisa.eu}), co-funded through the EU FP6
project RI-031513 and the FP7 project RI-222919, for support within the
DEISA Virtual Community Support Initiative.
JV was supported by Science and Technology Facilities Council (UK), and by the
Academy of Finland. TG acknowledges support from the Alexander von Humboldt Foundation.\\ 

\bibliography{Isocurvature}

\appendix

\section{On the sampling technique} \label{sec:sampling}

\subsection{Bayesian inference}
Bayesian statistics provides a good method to approach the two common
problems of parameter estimation and model comparison. It is based on
Bayes' theorem which states that for a set of parameters $\mathbf
{\Theta}$, in a model $M$, with data $\mathbf D$ it holds
\be
\mathcal{P}(\mathbf{\Theta}) = \frac{\mathcal{L}(\mathbf{\Theta})
\Pi(\mathbf{\Theta})}{\mathcal{Z} (M)},
\ee
where the posterior probability distribution of the parameters is
$\mathcal{P}(\mathbf{\Theta}) =
P(\mathbf{\Theta | \mathbf{D}, M}) $, the likelihood is $
\mathcal{L}(\mathbf{\Theta}) = P (\mathbf {D} | \mathbf {\Theta}, M)
$, the prior is $ \Pi(\mathbf{\Theta}) = P (\mathbf {\Theta} | M) $
and the evidence is $ \mathcal{Z} (M) = P (\mathbf{D} | M) $.

When estimating parameters for a given model, the standard practice is
to ignore the evidence factor, and to estimate the posteriors using the
standard MCMC method.

For model selection the evidence is instead crucial, since the ratio of
evidences reflects the relative probabilities of the models. The evidence can be
computed by the integration over all the dimensions $D$ of the parameter space
\be \label{eq:zint}
\mathcal{Z} (M) = \int \mathcal{L}(\mathbf{\Theta})
\Pi(\mathbf{\Theta}) d^D \mathbf{\Theta}.
\ee
This expression incorporates automatically Occam's principle of simplicity by penalizing models with extra parameters.
When comparing two models $A$ and $B$, the important quantity is the logarithmic difference in the evidences, also known as Bayes factor: $\Delta \ln \mathcal{Z} = \ln \mathcal{Z} (A) - \ln \mathcal{Z} (B) $. Then the model selection is qualitatively achieved using Jeffreys scale, which states that $ \Delta \ln \mathcal{Z} < 1 $ is not significant, $ 1 < \Delta \ln \mathcal{Z} < 2.5 $ is significant,  $ 2.5 < \Delta \ln \mathcal{Z} < 5 $ is strong, and  $  \Delta \ln \mathcal{Z} > 5 $ is decisive. In the main text we call  $\ln \mathcal{Z}$ as $\ln(\mr{ev})$ for clarity.

The standard method of thermodynamic integration, which is generally
used to calculate the evidence, is very intensive and expensive, typically
requiring the
evaluation of the likelihoods for
$10^6$---$10^7$ models, and has been hindering the widespread use of Bayesian
model comparison.

\subsection {Nested sampling}

The aforementioned problems are conveniently solved by the nested sampling
method \cite{Skilling:2004}. In this technique, the  
integral of Eq.~(\ref{eq:zint}) is replaced by a simpler 1-d integral
\be \label{eq:1dint}
\mathcal{Z} (M) = \int_0^1 \mathcal{L}(X) dX,
\ee
where the new variable $X$ represents the \emph{prior volume},
identical to the parameter space volume in case of uniform priors, and is
defined by $dX = \Pi (\mathbf{\Theta}) d^D \mathbf{\Theta}$, i.e.
\be
X(\lambda) = \int_{\mathcal{L}(\mathbf{\Theta}) > \lambda}  \Pi
(\mathbf{\Theta}) d^D \mathbf{\Theta},
\ee
where the integration is over the region contained in the iso-likelihood contour
defined by $\lambda$.
Thus the problem of calculating the evidence is reduced to the
evaluation of the likelihoods $\mathcal{L}_j$ at a series of points of
decreasing value
$X_j$, so that the 1-d integration of Eq.~(\ref{eq:1dint}) can be
performed by summation as
\be
\mathcal{Z} (M) = \sum_{i=1}^{N_{\mathrm{max}}} \mathcal{L}_i w_i,
\ee
where the weights $w_i$ can be given e.g. by a simple trapezoidal rule.

In more detail, the sampling of the $X_j$ can start with a uniform
sampling of $N$ points (often called as 'live points') within the priors, and then works its way up
the likelihood
surface by discarding at each iteration the lowest likelihood point
and replacing it with a higher one.
The process  is terminated when some accuracy criterion is satisfied.

Once the evidence is known, the posteriors can be easily evaluated as
a byproduct by
using the set of  points discarded at each iteration, giving each
point a weight
\be
p_i = \frac { \mathcal{L}_i w_i}{\mathcal{Z}}.
\ee

\subsection {MultiNest}
After the conceptual introduction by
\cite{Skilling:2004}, this method has been first applied to cosmology
in a simple case
by \cite {Mukherjee:2005wg}. Its most sensitive part, the sampling
technique, has been subsequently greatly refined by \cite {Shaw:2007jj}
and \cite{Feroz:2007kg} to minimize the required number of likelihood
evaluations and to deal efficiently with a series of possible
pathologies, such as multi-modal posterior distributions and strongly
curved parameter degeneracies.

Finally, an even more robust and
efficient development has been released by \cite{Feroz:2008xx} for
applications in cosmology, astronomy and particle physics. The
package, available for public use from \href {http://www.mrao.cam.ac.uk/software/multinest} {http://www.mrao.cam.ac.uk/software/multinest}, contains an easily usable interface
for the CAMB/Cosmomc cosmology code \cite{CAMB,COSMOMC}. The user has simply to tune three
parameters: the tolerance (accuracy), the number of live points $N$, and the maximum efficiency
$e$, which sets how aggressively (or conservatively) we want the reduce
the parameter space at each iteration. Another very attractive feature
of this method is that any need of a proposal matrix for the
parameters' covariance, a well known hassle for MCMC users, is now
completely superseded.

The analysis presented in this paper would have been impossible
with the conventional MCMC method. 
With MultiNest, the curved cases, which were the toughest, took
$30\,000$ --- $70\,000\,$ CPUh each. The extensive
comparisons presented in this paper took a total of $\sim500\,000\,$CPUh, but
this was doable in large super computers,
since the MultiNest algorithm scaled linearly
(with MPI parallelization) up to $\gtrsim 100$ CPUs in our case and
CAMB, which produces the theoretical predictions, scaled well up to 
$4$ --- $8\,$ CPUs with openMP. An efficient configuration turned out to be
$\sim$ 100 MPI $\times$ 6 openMP threads in the main runs.

In most of the cases reported in this paper we set
the efficiency parameter in MultiNest to 0.3, the tolerance (accuracy)
parameter to 0.5, and the number of live points to $N = 400$ .

\label{lastpage}

\end{document}